\documentclass[fleqn,twoside]{article}
\usepackage{espcrc2}
\usepackage{graphicx,amsmath,amssymb,cite}

\begin{document}
%%%%%%%%%%%%%%%%%%%%%%%%%%%%%%%%%%%%%%%%%%%%%%%%%%%%%%%%%%%%%%%%%%%%%%%

\title{Lattice status of gluonia/glueballs}

%%%%%%%%%%%%%%%%%%%%%%%%%%%% NEW SWITCHES %%%%%%%%%%%%%%%%%%%%%%%%%%%%%%

\author{Craig~McNeile\address[LIV]{
Department of Physics and Astronomy\\
The Kelvin Building\\
University of Glasgow\\
Glasgow G12 8QQ\\
U.K
}
}

\begin{abstract}
I briefly review lattice QCD calculations that study
the $0^{++}$ glueball and discuss implications for light flavour singlet
$0^{++}$ mesons.
%%%
\end{abstract}

% typeset front matter (including abstract)
\maketitle

\section{Introduction}  \label{se:section}

The $0^{++}$ glueball in pure gauge theory is a well defined
quantity that can be used to test our understanding and
ability to solve theory. It is also
interesting to understand how, or if, the $0^{++}$
glueball contributes to the physical 
light flavour singlet $0^{++}$ mesons.

Finding glueball degrees of freedom in the 
flavour singlet mesons is complicated, because
other non-perturbative objects such as 
tetraquark,
meson molecule or, 
even quark-antiquark degrees of freedom
can also be building blocks of scalar mesons. 
There are reviews that discuss these broader issues in
more detail~\cite{Liu:2007hm,McNeile:2007fu,Prelovsek:2008qu}.

\subsection{Background to lattice QCD}

Lattice QCD is based on a Monte Carlo process where 
a statistical sample of vacuum gauge fields is produced.
On each sample of the QCD vacuum, an
interpolating operator creates a hadron and after a
specific time interval the hadron is destroyed. The choice of
interpolating operator is particularly important for hadrons where it
is not clear how the hadron is built out of quarks and
gluons.

For example, to create a light flavour singlet $0^{++}$ hadron,
possible interpolating operators are
\begin{eqnarray}
O_1 & = &  \overline{q}q  \label{eq:standard} \\
O_2 & = &  \overline{q}\gamma_5 q \overline{q} \gamma_5 q
\label{eq:hybrid} \\
O_3 & = & U_{plaq}
\label{eq:glueball}
\end{eqnarray}
where $U_{plaq}$ is a spatial 
plaquette of gauge fields
with $0^{++}$ symmetry, and 
$q$ is a light quark operator. 

The majority of recent lattice QCD calculations
that include the dynamics of sea quarks have pion masses as low as 300
MeV~\cite{McNeile:2007fu}, with a range of volumes and lattice
spacings. These parameters have allowed lattice QCD to make
contact with chiral perturbation theory for light
pseudoscalar mesons~\cite{Leutwyler:2008ma}.
For some quantities, there are now precision results
from lattice QCD. For example the HPQCD collaboration
obtained $m_c(m_c)=1.268(9)$ GeV
for the mass of the charm quark~\cite{Allison:2008xk}.
The LHPC collaboration~\cite{Edwards:2005ym} computed the 
nucleon axial charge to be $g_A$ = 1.212(84)
from unquenched lattice QCD. The low lying glueball spectrum in
pure $SU(3)$ gauge theory was accurately computed
nearly ten years ago~\cite{Morningstar:1999rf,Chen:2005mg}

Unfortunately, lattice QCD calculations of scalar mesons 
are not as accurate
as those of other quantities.
The lattice QCD correlators for scalar
mesons are more noisy than for $\rho$ and $\pi$ mesons, so much higher
statistics are required.  The light scalar mesons decay via S-wave
decays, and current lattice QCD calculations are in the quark mass
regime where some decay channels to two mesons are open.  The results
I will present for flavour singlet $0^{++}$ mesons largely use the
last generation of lattice QCD calculations that are quenched,
 or dynamical
QCD calculations with pion masses 
above 500 MeV~\cite{Allton:2001sk,AliKhan:2001tx}.

\subsection{Glueballs as a theoretical testing ground}

The masses of glueballs from quenched QCD are a good test of calculational
techniques, such as methods based on a gauge/gravity 
duality (see~\cite{Caselle:2000tn} for a review).
Many of these new methods produce results in the large $N$ (number of colors) limit.
Teper and collaborators~\cite{Lucini:2001ej}
have studied glueballs for different $SU(N)$
groups on the lattice. In general they find
the $\frac{1}{N^2}$ corrections to the glueball masses 
in the large $N$ limit
to be small.
For example~\cite{Lucini:2001ej} 
the mass of the $m_{0^{++}}$ glueball
in units of string tension $\sigma$ was found to be 
\begin{equation}
\frac{m_{0^{++}}}{ \sqrt{ \sigma}} = 3.37(15) + \frac{1.93(85)}{N^2}
\end{equation}

Some of the methods based on gauge/gravity duality 
produce results in the strong coupling limit, so it is (perhaps)
interesting to compare them with older lattice results computed 
using a
strong coupling expansion in $\frac{1}{g^2}$. It is not always
easy to get a physical scale from all calculations, so
I compare the dimensionless ratio of the mass of the
$0^{++}$ and $2^{++}$ glueballs in table~\ref{tb:glueballSTRONG}.
The black hole and Klebanov-Strassler results are based on
calculations using gauge/gravity that use different metrics.
\begin{table}[tb]
\centering
\begin{tabular}{|c|c|} \hline
Method & $m_{2++} / m_{0++}$ \\ \hline
lattice $SU(3)$~\cite{Morningstar:1999rf}   & 1.39(4)   \\ 
lattice $SU(N)$~\cite{Lucini:2001ej}        & 1.46(11)   \\ \hline
strong coupling lattice~\cite{Smit:1982fx}  & 1.2 - 1.25 \\
black hole~\cite{Brower:2000rp}             & 1.73 \\
Klebanov-Strassler~\cite{Amador:2004pz}     & 1.37 \\
Frasca, strong coupling~\cite{Frasca:2008tg}        & 1.33 \\
\hline
\end{tabular}
\caption{Glueball ratios from different calculations}
\label{tb:glueballSTRONG}
\end{table}
Considering that the method that uses the Klebanov-Strassler~\cite{Amador:2004pz}
metric has a massless $0^{++}$ state, and that the calculation 
by Frasca~\cite{Frasca:2008tg} in pure Yang-Mills has an additional $0^{++}$ state
around 530 MeV, the result from the strong coupling lattice QCD calculation
is in fair agreement with the SU(3) result.
Brower et al.~\cite{Brower:2000rp}
compare in detail their results from a
gauge/gravity duality calculation with those
from lattice QCD~\cite{Morningstar:1999rf}.

\subsection{Flavour singlet $0^{++}$ mesons}

In nature and unquenched lattice QCD calculations glueball
and $\overline{q}q$ operators will mix, so glueballs
do not exist as separate particles.
The lightest flavor singlet $0^{++}$ mesons
listed in the PDG~\cite{Yao:2006px} are:  
$f_0(600)$, $f_0$(980), $f_0$(1370), $f_0$(1500), and $f_0$(1710),
so it is expected that $0^{++}$ glueball degrees of freedom will
contribute to some of these mesons.
There are claims that the $f_0$(980) and $f_0(600)$ may be  molecules
or tetraquark~\cite{Yao:2006px}, so may not couple to
$\overline{q}q$ interpolating operators in lattice QCD calculations.

Morningstar and Peardon~\cite{Morningstar:1999rf}
obtained $M_{0^{++}}$ = 1730(50)(80) MeV
for the mass of the lightest $0^{++}$ glueball from
quenched QCD.
Chen et al. ~\cite{Chen:2005mg} recently
found $M_{0^{++}}$ = 1710(50)(80) MeV from quenched
QCD.
The quark model predicts that there should only
be two $0^{++}$ mesons between 1300 and 1800 MeV, so
if the mixing between the glueball and $\overline{q}q$
operators is weak, then the $0^{++}$ glueball is hidden inside
the $f_0(1370)$, $f_0(1500)$ and $f_0(1710)$ mesons.
Klempt and Zaitsev have recently argued that the 
experimental data don't support that the $f_0(1370)$ is a separate
resonance~\cite{Klempt:2007cp}.

Weingarten and Lee~\cite{Lee:1999kv} used quenched lattice QCD to
estimate the mixing matrix between the glue and $\overline{q}q$
states. Weingarten and Lee~\cite{Lee:1999kv} predicted that the
$f_0(1710)$ meson was 74(10)\% $0^{++}$ glueball, and hence the mixing
between the $0^{++}$ glueball and $\overline{q}q$ states is weak.

There are claims~\cite{Mennessier:2007wk} that continuum phenomenology
is more consistent with a sizable contributions from the $0^{++}$
glueball to the $f_0(600)$ and $f_0(980)$ mesons, so it is important 
to study the effect of sea quarks on the masses of the state created by glueball 
interpolating operators.

%%
%% history of unquenched
%%
The SESAM collaboration studied the glueball spectrum
on unquenched lattices~\cite{Bali:2000vr}.
McNeile and Michael studied the light $0^{++}$ spectrum
with unquenched QCD~\cite{McNeile:2000xx} at a coarse
lattice spacing
and found the mass of the lightest flavour singlet
$0^{++}$ meson was very light.
Using $0^{++}$ glueball operators, 
Hart and Teper~\cite{Hart:2001fp} found that 
\begin{equation}
M_{0^{++}UNquenched} = 0.85(3) M_{0^{++}Quenched}
\end{equation}
at a fixed lattice spacing of 0.1 fm.
The UKQCD collaboration~\cite{Gregory:2005yr}
separately studied $0^{++}$ glueball 
and  $0^{++}$ $\overline{q}q$
operators on improved staggered gauge configurations,
however higher statistics and an analysis similar to the one
by Bernard et al.~\cite{Bernard:2007qf}
is required.

Unfortunately, the existing unquenched lattice QCD calculations
of the flavour singlet $0^{++}$ mesons don't have the 
range of lattice spacings where a continuum extrapolation
can be attempted.
In quenched QCD it was found that the lattice spacing 
dependence of the mass  of the $0^{++}$ 
glueball was strong. 
The use of a Symanzik improved gauge action by 
Chen et al.~\cite{Chen:2005mg}
and, Morningstar and Peardon~\cite{Morningstar:1999rf},
produced a smaller dependence on the lattice spacing
of the scalar $0^{++}$ glueball mass, than for calculations
that used the Wilson plaquette action. This 
is relevant to unquenched calculations, because 
any suppression of the mass of the flavour singlet $0^{++}$
mass may be due to lattice spacing effects.

The SCALAR collaboration~\cite{Kunihiro:2003yj}, 
used unquenched lattice QCD,
with Wilson fermions and the Wilson gauge action,
to study the $0^{++}$ mesons.
At a single lattice spacing a $\sim$ 0.2 fm,
with $\overline{q}q$ interpolating operators
only, they obtain $m_{\overline{q}q} \sim $ $m_{\rho}$.
The lattice spacing dependence of this result needs
to quantified.

\begin{figure}
\centering
\includegraphics[%
  scale=0.3,
  angle=270,
  origin=c]{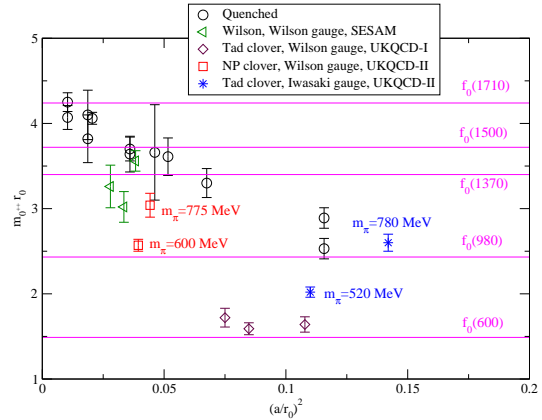}
\vspace{-2.5cm}
\caption{Summary of unquenched results for lightest flavour
singlet $0^{++}$ mesons from~\cite{Hart:2006ps}.
The unquenched results are from SESAM~\cite{Bali:2000vr},
UKQCD-I~\cite{McNeile:2000xx}, and UKQCD-II~\cite{Hart:2006ps}.}
\label{fig:unquenchGLUEBALL}
\end{figure}

In unquenched QCD, both glue and $\overline{q}q$ states
will couple to singlet $0^{++}$ mesons, so it is better
to do a variational fit with both types of operators as 
basis interpolating operators. 
The variational technique analysis of the 
singlet $0^{++}$ mesons was done by
Hart et al.~\cite{Hart:2006ps}.
A combined fit
to $0^{++}$ glue and $\overline{q}q$ interpolating
operators with two types of spatial smearing sources  was 
done. The calculation used the non-perturbative improved
clover action at a single lattice spacing~\cite{Allton:2001sk}.
Configurations from CP-PACS~\cite{AliKhan:2001tx}
with the Iwasaki gauge
action and tadpole improved clover action were also used 
in the analysis, because this calculation should be
less affected by lattice artifacts.
A summary plot of the results, in units of $r_0$ ($1/r_0 \sim$ 400 MeV) 
is in figure~\ref{fig:unquenchGLUEBALL}
(updated from~\cite{Hart:2006ps}).
%%%
%%%%%%%%%%%%%%%%%%%%%%%%%%%%%%%%%%
The data with the bursts and squares (with the pion masses
written near them)
in figure~\ref{fig:unquenchGLUEBALL}  
shows an additional reduction of the mass of the $0^{++}$ state over 
the pure glueball operators, as used by Hart and 
Teper~\cite{Hart:2001fp}. 

Mathur at al.~\cite{Mathur:2006bs} recently
claimed to get a result for the mass of the
$f_0(600)$ ($\sigma$)
from quenched lattice QCD with pion masses
as low as 180 MeV. Using the
interpolating operator in equation~\ref{eq:hybrid},
they obtain \mbox{$m_{f_0(600)} \sim 550$ MeV}. 
Mathur et al.'s~\cite{Mathur:2004jr} calculation is discussed in 
slightly more detail in~\cite{McNeile:2007fu}. The effect of 
sea quarks on this calculation needs to be quantified.

In~\cite{Hart:2006ps} an attempt was made to compute the 
decay width for $f_0$ decay to two pions. Unfortunately 
much higher statistics will be required to obtain an accurate
value for that width.
Recent unquenched lattice QCD calculations 
have light enough quarks
that the two meson decays of some scalar mesons 
are allowed and some preliminary evidence for $0^{++}$
decay has been presented~\cite{Michael:2007vn}.

Perhaps a more mundane issue with the improving unquenched lattice
QCD calculations of flavour singlet quantities is just increasing the 
statistics in the Monte Carlo estimate. In table~\ref{se:statsglueball},
I show the number of measurements done in some quenched and unquenched
calculations of $0^{++}$ glueballs. This type of comparison between 
number of estimates, can be misleading because it 
depends on the autocorrelation times. The qualitative message from
table~\ref{se:statsglueball} is that unquenched calculation of
the flavour singlet $0^{++}$ mesons need at least 10 times
as much statistics as currently used. Lattice QCD calculations with
higher statistics are definitely required to study unquenching
on the mass obtained from $2^{++}$ glueball interpolating 
operator~\cite{Hart:2001fp}.
There is a high statistics unquenched lattice QCD calculation in progress
of light flavour singlet mesons,
that has generated 6000 configurations at a single value of the lattice 
spacing~\cite{Gregory:2005yr,Gregory:2007ce}. 
\begin{table}[tb]
\centering
\begin{tabular}{|c|c|c|} \hline
Group & Method & statistics \\ \hline
Morningstar et al.~\cite{Morningstar:1999rf} &  quenched &  6360 \\
Chen et al.~\cite{Chen:2005mg}  & quenched  &  10000 \\
Hart et al.~\cite{Hart:2006ps}  & unquenched &   500  \\
\hline
\end{tabular}
\caption{Comparison of statistics between quenched and unquenched glueball calculations.}
\label{se:statsglueball}
\end{table}

Chen et al.~\cite{Chen:2005mg} 
and Meyer~\cite{Meyer:2008tr} have recently computed 
the matrix element of the energy momentum tensor with
glueball states in quenched QCD calculations. 
Meyer~\cite{Meyer:2008tr} uses the 
computed matrix elements to study $J/\psi$ radiative decay.

\section{Conclusions}

There is ``some'' evidence that the flavour singlet $0^{++}$ 
$\overline{q}q$ and glueball
interpolating operators,
in unquenched lattice QCD calculations, are coupling to states around
or below 1 GeV~\cite{Hart:2006ps}. Although a continuum
extrapolation is required for definite results.
Unquenched lattice QCD calculations with
$0^{++}$ tetraquark interpolating operators are required
to clarify the composition of the lightest $0^{++}$ resonance.

%%\bibliographystyle{h-elsevier2}
%%\bibliography{latt05}

\end{document}